\documentclass[12pt]{iopart}
\begin{document}
\paper{Statistical correlations in a Coulomb gas with a test
charge}

\author{Henning Schomerus}
\address{Department of Physics, Lancaster University, Lancaster, LA1 4YB, UK}
\begin{abstract}
A recent paper [Jokela {\em et al.}, arxiv:0806.1491 (2008)] contains a
surmise about an expectation value in a Coulomb gas which interacts with an
additional charge $\xi$ that sits at a fixed position. Here I demonstrate the
validity of the surmised expression and extend it to a certain class of
higher cumulants. The calculation is based on the analogy to statistical
averages in the circular unitary ensemble of random-matrix theory and
exploits properties of orthogonal polynomials on the unit circle.
 \end{abstract}
\pacs{02.50.Sk, 05.20.-y, 05.30.Fk, 05.45.Mt} 

\section{Purpose and result} In a recent paper Jokela, J{\"a}rvinen
and Keski-Vakkuri studied $n$-point functions in timelike boundary
Liouville theory via the analogy to a Coulomb gas on a unit circle
\cite{jokela}. In this analogy, $N$ unit charges at position $t_i$
interact with additional charges of integer value $\xi_a$,
situated at position $\tau_a$. To illustrate this technique the
authors of \cite{jokela} considered the canonical expectation
value
\begin{equation}
\langle\cdot\rangle\equiv\frac{1}{Z}
\int\prod_{i=1}^N\frac{dt_i}{2\pi}\prod_{i<j}\left|\rme^{\rmi t_i}-\rme^{\rmi t_j}\right|^2
\prod_{i}\left|\rme^{\rmi \tau}-\rme^{\rmi t_i}\right|^{2\xi} (\cdot)
\end{equation}
(where $Z$ is a normalization factor so that $\langle 1 \rangle=1$) and
surmised that
\begin{equation}
\langle {\rm Re}\, a_1\rangle \equiv\left\langle\sum_{i}\cos(\tau-t_{i})\right\rangle=-\frac{\xi N}{N+\xi}.
\label{eq:surmise}
\end{equation}

In this communication I demonstrate the validity of
(\ref{eq:surmise}), and also compute expectation values of the
more general quantities
\begin{equation}
a_n\equiv \sum_{i_1<i_2<\ldots<i_n}\exp\left(\rmi \sum_{k=1}^n(t_{i_k}-\tau)\right).
\label{eq:an}
\end{equation}
As a result, I find
\begin{equation}
\langle a_n\rangle
=(-1)^n\frac{(N-n+1)^{(\xi)}(n+1)^{(\xi-1)}}{(N+1)^{(\xi)}(1)^{(\xi-1)}}
\quad \forall\,\, n=0,1,2,\ldots,N;\,\, \xi \geq 0,
\label{eq:result}
\end{equation}
where $(x)^{(y)}=\Gamma(x+y)/\Gamma(x)$ is the generalized rising
factorial (Pochhammer symbol).
 In
particular, the validity of (\ref{eq:surmise}) follows from
(\ref{eq:result}) by setting $n=1$.

Expression (\ref{eq:result}) will be obtained by relating the generating
polynomial
\begin{equation} \varphi_{N,\xi}(\lambda)\equiv\sum_{n=0}^N \langle a_n\rangle
(-\lambda)^{N-n}
\label{eq:pol}
 \end{equation}
to a weighted average of the secular polynomial in the circular
unitary ensemble (CUE). This in turn establishes a relation to the
Szeg\H{o} polynomial of a Toeplitz matrix composed of binomial
coefficients. This calculation sidesteps Jack polynomials and
generalized Selberg integrals, which can be used to tackle general
expectation values in  multicomponent Coulomb gases
\cite{forrester1}.

\section{Reformulation in terms of random matrices} The CUE is composed of
$N\times N$ dimensional unitary matrices $U$ distributed according
to the Haar measure. Identify $t_i$ with the eigenphases of such a
matrix. The joint probability distribution is then given by
\cite{mehta}
\begin{equation}
P(\{t_i\}_{i=1}^N)=z
\prod_{i<j}\left|\rme^{\rmi t_i}-\rme^{\rmi t_j}\right|^2,
\end{equation}
where $z$ is again a normalization constant. This expression can also be
written as the product of two Vandermonde determinants $\det V^+\det V^-$
with matrices $V_{lm}^{\sigma}=\rme^{\rmi \sigma(m-1) t_l}$. Furthermore, we
can write
\begin{equation}
\prod_i\left|\rme^{\rmi \tau}-\rme^{\rmi t_i}\right|^{2\xi} =[\det(1-U\rme^{-\rmi \tau})\det(1-U^\dagger \rme^{\rmi \tau})]^{\xi}.
\end{equation}
Finally, the expressions $a_n$ in (\ref{eq:an}) arise as the
expansion coefficients of the secular polynomial
\begin{equation}
\det(U\rme^{-\rmi \tau}-\lambda)=\sum_{n=0}^N a_n(-\lambda)^{N-n} .
\end{equation}
Note that in all these expressions $\tau$ can be shifted to any fixed value
by a uniform shift of all $t_i$'s, which leaves the unitary ensemble
invariant. Therefore the expectation values are independent of $\tau$.
Collecting all results, we have the identity
\begin{equation}
\varphi_{N,\xi}(\lambda)=\frac{\left\langle[\det(1-U)\det(1-U^\dagger)]^{\xi}\det(U-\lambda)\right\rangle_{\rm CUE}}
{\left\langle[\det(1-U)\det(1-U^\dagger)]^{\xi}\right\rangle_{\rm CUE}}.
\label{eq:sec}
\end{equation}
This can be interpreted as a weighted average of the secular polynomial in
the CUE.

\section{Random-matrix average}  Statistical properties of the secular
polynomial without the weight factor ($\xi=0$) have been
considered in \cite{haake}. Clearly, $\varphi_{N,0}=(-\lambda)^N$,
so that in this case the attention quickly moves on to higher
moments of the $a_n$. The main technical observation in
\cite{haake} which allows to address the case of finite $\xi$
concerns averages of expressions $g(\{t_i\}_{i=1}^N)$ that are
completely symmetric in all eigenphases. In this situation the
average can be found via
\begin{equation}
\langle g(\{t_i\}_{i=1}^N)\rangle_{\rm CUE}=\int\prod_i\frac{\rmd t_i}{2\pi} g(\{t_l\}_{l=1}^N) \det W,
\label{eq:cuesimp}
\end{equation}
where $W_{lm}=\rme^{\rmi t_m (l-m)}$. Equation (\ref{eq:cuesimp})  is simpler
than the general expression involving the product of two Vandermonde
matrices, since each eigenphase only appears in a single column of $W$.

In the present problem, the numerator in (\ref{eq:sec}) is
represented by the completely symmetric function
\begin{equation}
g_1(\{t_i\}_{i=1}^N)=\prod_{i=1}^N[(\rme^{\rmi t_i}-\lambda)(1-\rme^{\rmi t_i})^\xi(1-\rme^{-\rmi t_i})^\xi],
\end{equation}
while for the denominator we need to consider the similar expression
\begin{equation}
g_2(\{t_i\}_{i=1}^N)=\prod_{i=1}^N[(1-\rme^{\rmi t_i})^\xi(1-\rme^{-\rmi t_i})^\xi].
\end{equation}

Using the multilinearity of the determinant we can now pull each factor into
the $i$th column and perform the integrals. This delivers the representation
\begin{equation}
\varphi_{N,\xi}(\lambda)=\frac{\det(B-\lambda A)}{\det A},
\end{equation}
where the matrices $A_{lm}=(-1)^{l-m}{{2\xi}\choose{\xi+l-m}}$,
$B_{lm}=(-1)^{l-m+1}{{2\xi}\choose{\xi+l-m+1}}$ have entries given
by binomial coefficients. We now exploit the regular structure of
these matrices in two steps.

1) Matrix $B$ contains the same entries as matrix $A$, but shifted
to the left by one column index. In order to exploit this, let us
expand the determinant in the numerator into a sum of determinants
of matrices labeled by $X=(x_m)_{m=1}^N$, where we select each
column either from $A$ ($x_m={\rm A}$) or from $B$ ($x_m={\rm
B}$). [Note that we set these symbols in roman letters.] The
related structure of $A$ and $B$ then entails that $\det X$
vanishes if $X$ contains a subsequence $(x_m,x_{m+1})=({\rm
A},{\rm B})$. Consequently we only need to consider determinants
of matrices $X_n \equiv ({\rm B})_{m=1}^{n}\oplus({\rm
A})_{m=n+1}^N$, associated to sequences that contain $n$ leading
B's and $N-n$ trailing A's. As $A$ is multiplied by $-\lambda$,
$\det X_n$ contributes to order $(-\lambda)^{N-n}$. (Note that
$X_0=A$ and $X_N=B$.)

2) Next, consider the matrix $A_{N+1}$, where the subscript denotes the
dimension, and strike out the first row and the $n+1$st column
($n=0,1,2,\cdots,N$). This takes exactly the form of the matrix $X_{n}$ of
dimension $N$. Therefore, the expressions $(-1)^{n}\det X_{n}$ are the
cofactors of the first row of $A_{N+1}$. These, in turn, are  proportional to
the first column of $A_{N+1}^{-1}$, where the proportionality factor is given
by $\det A_{N+1}$. Consequently, taking care of all alternating signs,
\begin{equation}
\varphi_{N,\xi}(\lambda)= (-1)^N\frac{\det A_{N+1}}{\det A_N} \sum_{n=0}^N(A_{N+1}^{-1})_{1,1+n}\lambda^{N-n}.
\label{eq:aexp}
\end{equation}
Via steps 1) and 2) we have eliminated any reference to the matrix $B$.

\section{Orthogonal polynomials} Matrix $A$ is a Toeplitz matrix,
$A_{lm}=c_{l-m}$. In order to find the explicit expression
(\ref{eq:result}) we now make contact to the theory of orthogonal
polynomials on the unit circle \cite{barry}. Among its many
applications, this theory provides a general expression for the
inverse of any Toeplitz matrix in terms of Szeg\H{o} polynomials
$\psi_N(\lambda)$. For the case of real symmetric coefficients,
the inverse is generated via
\begin{equation}\fl
\frac{\lambda\mu^{N}\psi_N(\lambda)\psi_N(\mu^{-1})-\lambda^N\mu\psi_N(\lambda^{-1})
\psi_N(\mu)}{\lambda-\mu}
=  \frac{\det A_{N+1}}{\det A_N}\sum_{n,m=0}^N(A_{N+1}^{-1})_{m+1,n+1}\lambda^{N-n}\mu^{m}.
\label{eq:aexp1}
\end{equation}
Comparison of this equation with $m=0$ to (\ref{eq:aexp})
immediately leads to the identification of
$(-1)^N\varphi_{N,\xi}(\lambda)$ with the Szeg\H{o} polynomial
$\psi_N(\lambda)$ of degree $N$. These polynomials satisfy
recursion relations which for real symmetric coefficients take the
form \numparts
\begin{eqnarray}
\gamma_{N}=-\frac{1}{\delta_{N-1}} \oint\frac{\rmd\lambda}{2
\pi\rmi} \psi_{N-1}(\lambda) \sum_{n=-\infty}^\infty c_n\lambda^n
,
\\
\psi_{N}(\lambda)=\lambda\psi_{N-1}(\lambda)+\gamma_N\lambda^{N-1}\psi_{N-1}(\lambda^{-1}),
\\[.3cm]
 \delta_N=\delta_{N-1}(1-\gamma_N^2).
\end{eqnarray}
\endnumparts
The initial conditions are $\delta_0=c_{0}$, $\psi_0(\lambda)=1$. The numbers
$\gamma_N$ are known as the Schur or Verblunsky coefficients.

It can now be seen in an explicit if tedious  calculation that the
polynomials \numparts
\begin{eqnarray}
\psi_N(\lambda)=(-1)^N\varphi_{N,\xi}(\lambda)=
\sum_{n=0}^N\frac{(N-n+1)^{(\xi)}(n+1)^{(\xi-1)}}{(N+1)^{(\xi)}(1)^{(\xi-1)}}\lambda^{N-n}
\\=\lambda^N{}_2F_1(-N,\xi;-N-\xi;\lambda^{-1})
\end{eqnarray}
[with coefficients and expansion given in (\ref{eq:result}),
(\ref{eq:pol})] indeed fulfill the Szeg\H{o} recursion generated
by the binomial coefficients $c_n=(-1)^n{{2\xi}\choose \xi-n}$.
The recursion coefficients take the simple form \begin{equation}
\gamma_N=\frac{\xi}{\xi+N},\qquad
\delta_N=\frac{N!(2\xi+1)^{(N)}}{[(\xi+1)^{(N)}]^2}.
\end{equation}
\endnumparts
 This completes the proof of
(\ref{eq:result}), and also entails the validity of (\ref{eq:surmise}).

\ack

The author wishes to thank Niko Jokela, Matti J{\"a}rvinen and
Esko Keski-Vakkuri for useful correspondence regarding reference
\cite{jokela}, and Mitsuhiro Arikawa and Peter Forrester for
useful communications about reference \cite{forrester1}. This work
was supported by the European Commission, Marie Curie Excellence
Grant MEXT-CT-2005-023778 (Nanoelectrophotonics).

\end{document}